\documentclass[10pt,oneside,twocolumn,a4paper]{article}
\pdfoutput=1
\usepackage{graphicx}
\usepackage[utf8]{inputenc}
\usepackage{mathptmx}
\usepackage{tabularx}
\usepackage{ragged2e}
\usepackage[singlelinecheck=false]{caption}
\usepackage[T1]{fontenc}
\usepackage[numbers]{natbib}
\usepackage{amsmath}
\RequirePackage[blocks]{authblk}
\usepackage[english]{babel}
\usepackage{amsmath,amssymb,amsfonts}
\usepackage{graphicx}
\usepackage{xcolor}
\usepackage{footnote}
\usepackage{hyperref}
\usepackage{booktabs}
\usepackage{balance}
\def\BibTeX{{\rm B\kern-.05em{\sc i\kern-.025em b}\kern-.08em  T\kern-.1667em\lower.7ex\hbox{E}\kern-.125emX}}

\newcommand{\cities}{Berlin, Hamburg, Munich, Würzburg, Hildesheim, Cottbus, and Gütersloh}

\makeatletter
\let\ps@plain\ps@empty
\def\@xivpt{14bp}

\def\@sect#1#2#3#4#5#6[#7]#8{%
  \ifnum #2>\c@secnumdepth
    \let\@svsec\@empty
  \else
    \refstepcounter{#1}%
    \protected@edef\@svsec{%
      \ifnum #2<4
        \hb@xt@10mm{\csname the#1\endcsname}\relax
      \else
        \hb@xt@12mm{\csname the#1\endcsname}\relax
      \fi}%
  \fi
  \@tempskipa #5\relax
  \ifdim \@tempskipa>\z@
    \begingroup
      #6{%
        \@hangfrom{\hskip #3\relax\@svsec}%
          \interlinepenalty \@M #8\@@par}%
    \endgroup
    \csname #1mark\endcsname{#7}%
    \addcontentsline{toc}{#1}{%
      \ifnum #2>\c@secnumdepth \else
        \protect\numberline{\csname the#1\endcsname}%
      \fi
      #7}%
  \else
    \def\@svsechd{%
      #6{\hskip #3\relax
      \@svsec #8}%
      \csname #1mark\endcsname{#7}%
      \addcontentsline{toc}{#1}{%
        \ifnum #2>\c@secnumdepth \else
          \protect\numberline{\csname the#1\endcsname}%
        \fi
        #7}}%
  \fi
  \@xsect{#5}}
\renewcommand\LARGE{\@setfontsize\LARGE{16}{20}}
\def\abstract#1{\def\@abstract{#1}}
\def\abstractEn#1{\def\@abstractEn{#1}}
\def\titleEn#1{\def\@titleEn{#1}}
\def\@maketitle{%
  \newpage
  \null
  \let \footnote \thanks
    {\LARGE\bfseries\RaggedRight \@titleEn \par}%
    \vskip 1\baselineskip%
    {\normalsize
      \@author\par}%
    \vskip 2\baselineskip%
    \vskip \baselineskip%
    {\section*{Abstract}
      \@abstractEn}%
  \par
  \vskip 3\baselineskip}

\renewcommand\section{\@startsection {section}{1}{\z@}%
                                   {-3.5ex \@plus -1ex \@minus -.2ex}%
                                   {\baselineskip}%
                                   {\normalfont\Large\bfseries\RaggedRight}}
\renewcommand\subsection{\@startsection{subsection}{2}{\z@}%
                                     {\baselineskip}%
                                     {1ex}%
                                     {\normalfont\large\bfseries\RaggedRight}}
\renewcommand\subsubsection{\@startsection{subsubsection}{3}{\z@}%
                                     {1\baselineskip}%
                                     {3bp}%
                                     {\normalfont\normalsize\bfseries\RaggedRight}}
\renewcommand\paragraph{\@startsection{paragraph}{4}{\z@}%
                                    {1\baselineskip\@plus1ex \@minus.2ex}%
                                    {3bp}%
                                    {\normalfont\normalsize\RaggedRight}}
\renewcommand\subparagraph{\@startsection{subparagraph}{5}{\parindent}%
                                       {3.25ex \@plus1ex \@minus .2ex}%
                                       {-1em}%
                                      {\normalfont\normalsize\bfseries\RaggedRight}}
\parindent\p@
\makeatother
\DeclareCaptionLabelSeparator{enskip}{\enskip}
\title{Crowdsourced Network Measurements in Germany: Mobile Internet Experience from End User Perspective}
\titleEn{Crowdsourced Network Measurements in Germany: Mobile Internet Experience from End User Perspective}
\author{Anika Schwind$^*$, Florian Wamser$^*$, Tobias Hoßfeld$^*$, Stefan Wunderer$^+$, Erik Tarnvik$^\diamond$, Andy Hall$^\diamond$}
\affil{
    $^*$ \textit{University of W\"urzburg, Insitute of Computer Science}, W\"urzburg, Germany\\
  \{anika.schwind$|$florian.wamser$|$tobias.hossfeld\}@informatik.uni-wuerzburg.de\\
  $^+$ \textit{Nokia Networks}, Ulm, Germany, stefan.wunderer@nokia.com\\
  $^\diamond$ \textit{Tutela Ltd}, Victoria, Canada, 
  \{et$|$ahall\}@tutela.com
 }

\abstractEn{
Collecting and analyzing meaningful data in mobile networks is the key to assessing network performance. 
Crowdsourced Network Measurements (CNMs) provide insights beyond the network layer and offer performance and other measurements at the application and user-level towards Quality of Experience (QoE). 
In this paper, the mobile Internet experience for Germany is evaluated with the help of crowdsourcing from the perspective of an end user. 
We statistically analyze a dataset with throughput measurements on the end device from Tutela Ltd., which covers more than 2.5 million throughput tests across Germany from January to July 2019. 
We give insights into this emerging methodology and highlight the benefits of this method. 
The paper contains statistics and conclusions for several large cities as well as regions in Germany compared to general statements for Germany, since individual measurements and averages often only imprecisely reflect the situation. 
The goal is to give a holistic view of the performance of the current mobile network in Germany. 
Reading this paper, it becomes evident that reliable statements about the quality of the mobile network for Germany depend on a large number of peculiarities in different regions with their own performance characteristics due to different network deployments and population numbers.
}

\begin{document}

\maketitle

\section{Introduction}\label{sec:introduction}

The collection and analysis of meaningful data has always been the key to optimizing the performance of communication networks. 
Previously, operators or providers have captured and analyzed network data and Quality of Service (QoS) factors at the network level to assess the network perspective. Using Crowdsourced Network Measurements (CNMs), it is possible to get insights beyond the network layer, i.e. on application layer and user level towards Quality of Experience (QoE).

CNMs allow network operators to think outside the box using the mass of end user devices for gaining measurement data. 
It gives a much better holistic understanding of the impact of network challenges or issues on the quality experienced by end users. 
The ultimate goal is to use CNMs combined with other network and user data to improve QoE for applications, but also for regulatory entities to identify and predict coverage, network, and quality issues. Thus, in this paper, we will show how CNM data can be used to evaluate the QoS and QoE of end users in Germany by evaluating the measured throughput and compare it among different states and cities. Thereby, we will utilize an existing crowdsourced measurement dataset of Tutela Ltd., covering more than 2.5 million throughput tests in Germany from January until July 2019.

In our paper we will highlight the benefits of going beyond pure speed-tests by utilizing crowd data.
After discussing related work as well as related measurement approaches, we will go into detail how Tutela performs its measurements and how these measurements can be interpreted.
After that, we present some exemplary evaluations and show that CNMs can help to get insights into the user perceived Mobile Broadband (MBB) experience.
Not only the individual measurements play a role here, but rather the overall statement and the understanding of the validity of an individual value as an overall statement is in focus, which results in a differentiated view of the measurements at different network locations and regions. 
Thus, the results of the paper are covering detailed throughput statistics for Germany as well as results for the 16 German states in comparison of some highly as well as sparsely populated areas.
We will show that, for example, Hamburg leads the ranking of the average download throughput per state with about 18.1\,Mbps, while Saarland is at the low end, having an average of 12.3\,Mbps. 
The paper shows that the mean throughput as well as the variance differs between regions.
In addition, we map the results to QoE related performance indicators to include the user perspective of the network.

The remainder of this work is structured as follows. 
Section~\ref{sec:relWork} provides background and discusses related work as well as available network reports.
The methodology to measure the network speed is described in Section~\ref{sec:methodology}. 
Section~\ref{sec:evaluation} deals with the measurement data and compares the results of the German states as well as of different German cities. 
Finally, Section~\ref{sec:conclusion} concludes the paper. 
\section{Background on Crowdsourcing and Related Work}\label{sec:relWork}

Crowdsourcing is the methodology of processing a task by a large group of people instead of a designated agent~\cite{howe2006rise}.
Best practices and recommendations for crowdsourced QoE assessment are discovered in~\cite{hossfeld2013best, hossfeld2014best}.
For network measurements, crowdsourcing has three major advantages: it makes it much easier to cover a wide range of situations and users, it allows entities other than the network operator to assess the performance and other characteristics of a network, independently, with a coverage that is not feasible using other methods such as drive testing, and it offers the possibility to collect statistics from end user perspective. 
A comparison of crowdsourcing with traditional measurement techniques and best practices how to design crowdsourced network measurements issues was done in~\cite{hirth2015crowdsourced}. 
There are two ways of doing crowdsourcing studies: either workers are paid to process a task or applications on the end user's smartphone are used to collect key performance indicators (KPIs).
With the second way, crowdsourced network measurements can be seen as a special case of crowdsensing where user devices act as environmental sensors. 
Crowdsourced measurement data (crowd data) offers new possibilities and can be used for various applications such as the benchmarking of network operators, providers, technologies, or countries as well as e.g., be used for monitoring, planning, and optimization of the network. 
The ultimate goal is to use crowd data -- combined with other network and user data -- to improve QoE, but also for regulatory purposes, e.g. to identify issues with coverage, or network.

In recent years, crowdsourced network measurements become increasingly relevant in research.
One example of this is drive testing.
While drive testing is a commonly used tool to characterize mobile networks, there is further development towards using a large community and war-driving to discover WiFi access points and draw a map for coverage~\cite{kim2006risks}.
There are a number of algorithms available for crowdsourced measurement-based cell tower localization to create a coverage map, e.g.,~\cite{neidhardt2013estimating,li2017identifying}.
The coverage map accuracy from the perspective of device diversity is investigated in~\cite{fida2018impact}.
Furthermore, there are also other ways of measuring and comparing Internet service providers (ISPs) using crowdsourced data.
In~\cite{Bischof2011}, the authors used crowd data collected from peer-to-peer BitTorrent users to compare the performance of ISPs from end user perspective.
Crowdsourced network measurements can also be done by collecting information during the use of smartphone applications.
Here, different KPIs can be measured on several layers, from context parameters like cultural background through network parameters like signal strength up to application parameters like number of stalling and user-focused parameters like browser session time.
Especially video streaming applications are well used options to collect crowd data on the smartphone of the end users~\cite{wamser2015yomoapp, wassermann2019on, schwind2019dissecting, schwind2019mobile}.

Due to this development, the number of crowdsourced network measurement service providers increased in the last few years, for example, Tutela\footnote{Tutela Technologies, \url{https://www.tutela.com/}}, 
Ookla\footnote{Ookla, \url{https://www.speedtest.net/}}, 
P3\footnote{P3 communications, \url{http://p3-networkanalytics.com/}}, 
QoSi\footnote{QoSi, \url{https://www.5gmark.com}}, and Opensignal\footnote{Opensignal, \url{https://www.opensignal.com/}}, which use the smartphone of the end users as measurement device.
These companies regularly publish reports on the mobile network experience in Germany~\cite{4gmark2017germany,tutela2019germany,p3communications2019germany,opensignal2019germany}, in which they compare network operators, coverage, and speed of their networks.

Nevertheless, to the best of our knowledge, there is as yet no comprehensive evaluation of the broadband coverage and connection speed in Germany, comparing the performance of states and cities on a national level.
\section{Methodology}\label{sec:methodology}

The dataset used belongs to Tutela Ltd., an independent crowdsourced data company with a global panel of over 300 million smartphone users. Tutela collects data and conducts network tests through software embedded in a variety of over 3\,000 consumer applications. Although started at random times, measurements are performed in the background in regular intervals if the user is inactive, and information about the status of the device and the activity of the network and the operating system are collected. The data is correlated, grouped and evaluated according to device and network status (power saving mode, 2G/3G/4G connectivity). Tests are conducted against the same content delivery network. 

Tutela measures the network quality based on the real performance of the actual network user, including situations when a network is congested, or the user is throttled by tariff. The results in this paper are based on throughput testing in which 2\,MB files are downloaded via Hypertext Transfer Protocol Secure (HTTPS). The size reflects the median of the web page size on the Internet.

\subsection{Dataset}

Within the used dataset for Germany, 3\,891\,839 crowdsourced network measurements are included. 
After filtering incomplete entries, 2\,555\,557 measurements remain.
Data was collected during half a year from January 2019 to July 2019. 
In addition to meta information like date, geo-coordinates, region and city, the dataset includes information on current network performance, including upload throughput, download throughput, average latency, average jitter, and percentage of packet loss, as well as details about the used network connection (such as technology and frequency band) as well as the
device (such as OS and device make and model) are collected.

\begin{figure} [tbh]
    \centering
    \includegraphics[width=.55\columnwidth]{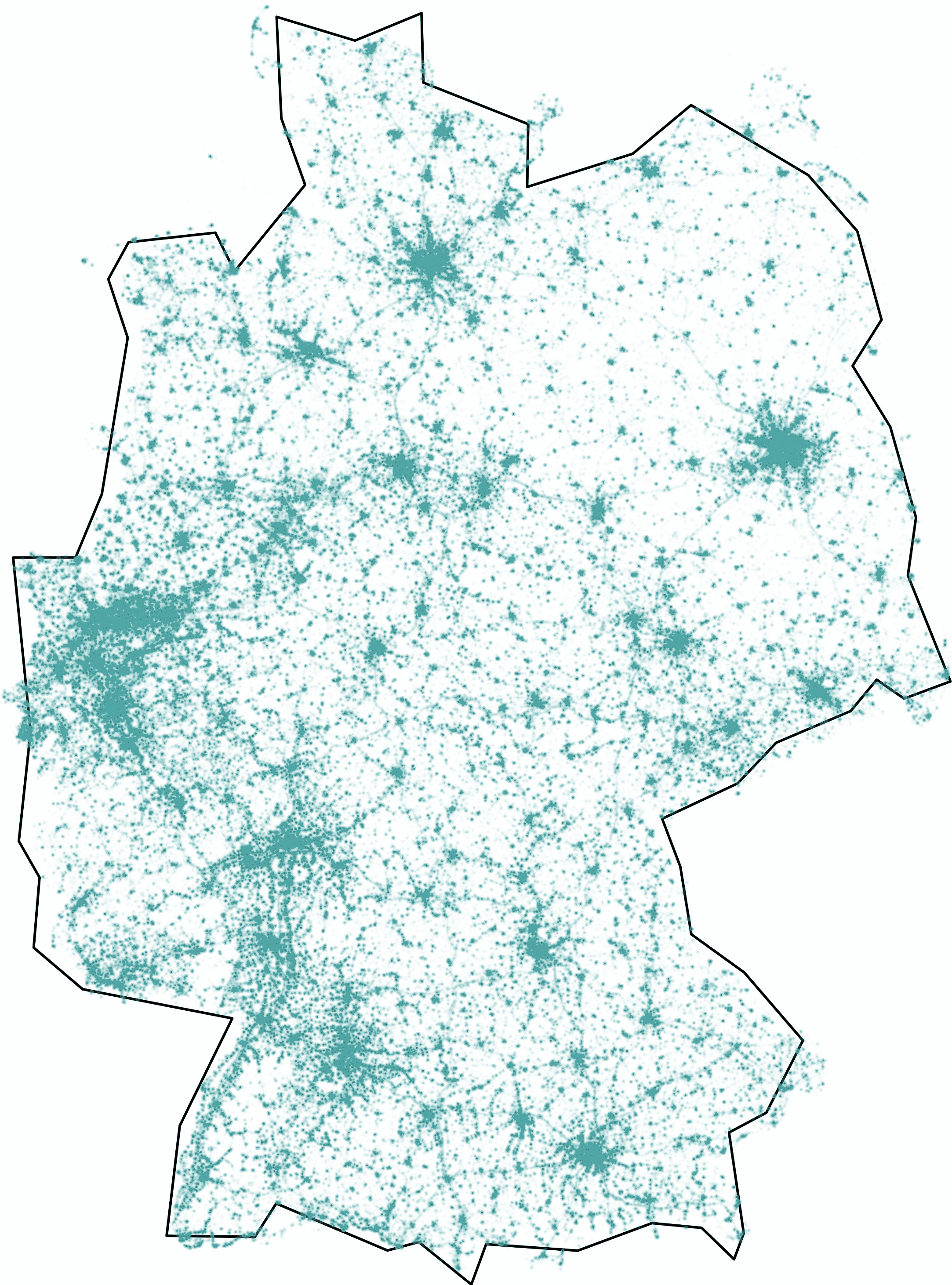}
   \caption{Distribution of all measurements in Germany.}
      \label{fig:map_measurements} 
\end{figure}

\textbf{Figure~\ref{fig:map_measurements}} shows the distribution of the measurements on the map.
As the data is collected using crowdsourcing, measuring points are not distributed evenly, but are often to be seen in urban centers and cities.
The mean number of measurements per square kilometer is about 7\,150. Having a look at the distances between the measurements, the mean distance to the nearest neighbor is 26.2\,m with a maximum of 9\,941.89\,m and a standard variation of 72.9\,m.

\begin{figure} [t]
    \centering
    \includegraphics[width=\columnwidth]{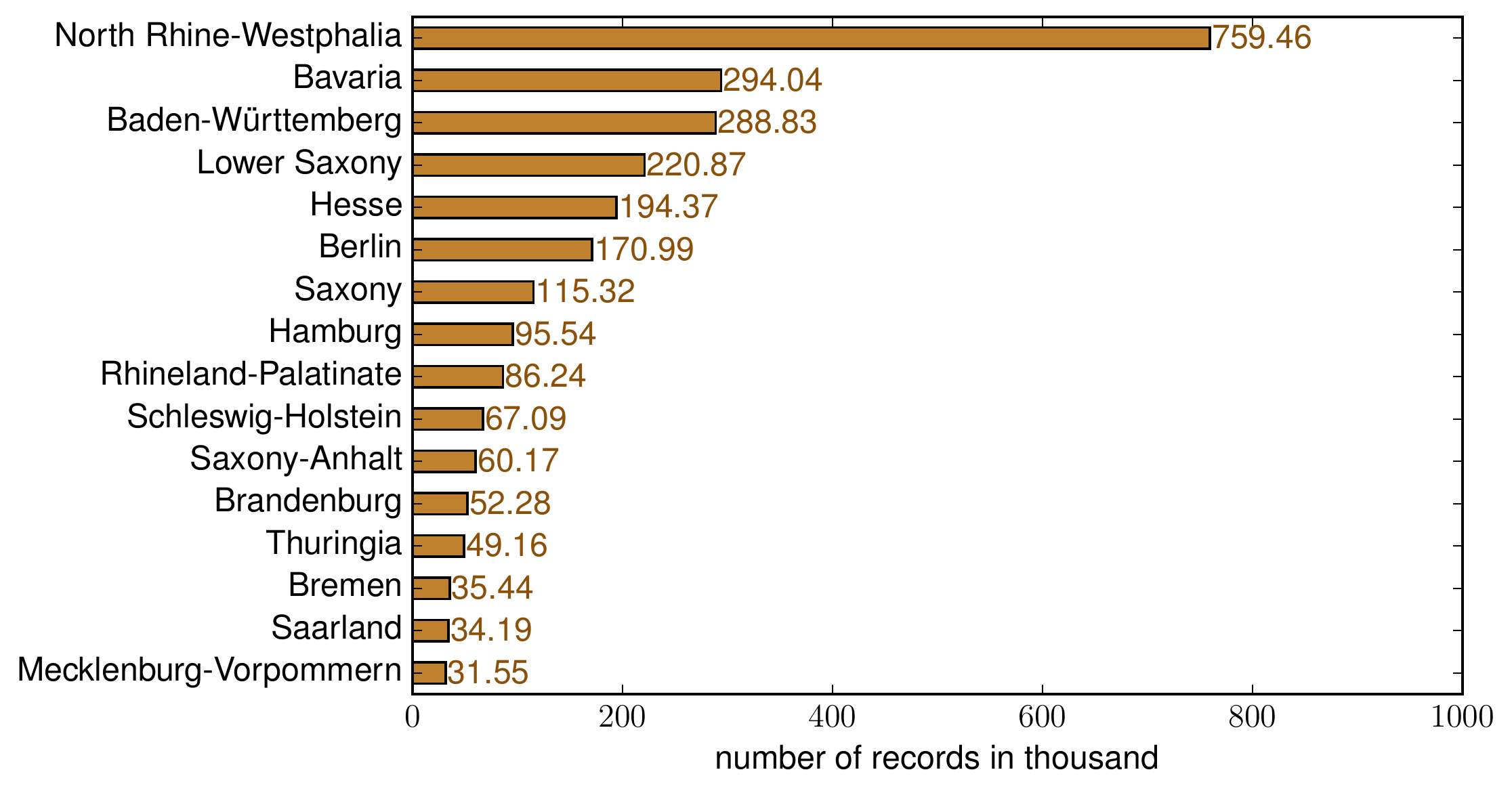}
  \caption{Number of measurements per state.}
      \label{fig:regionNumMeasurements} 
\end{figure}

\textbf{Figure~\ref{fig:regionNumMeasurements}} shows the distribution of the number of measurements over the 16 German states.
The highest number of measurements can be seen in North Rhine-Westphalia (759\,465 measurements) while the lowest number was measured in Mecklenburg-Vorpommern (31\,552 measurements).
Due to the fact that we used crowdsourced measurements, a high linear correlation (Pearson correlation: $\tau=0.94$, $p{\text-}value\leq0.01$) between the number of data points per state and the number of inhabitants is visible.

\begin{figure} [t]
    \centering
    \includegraphics[width=\columnwidth]{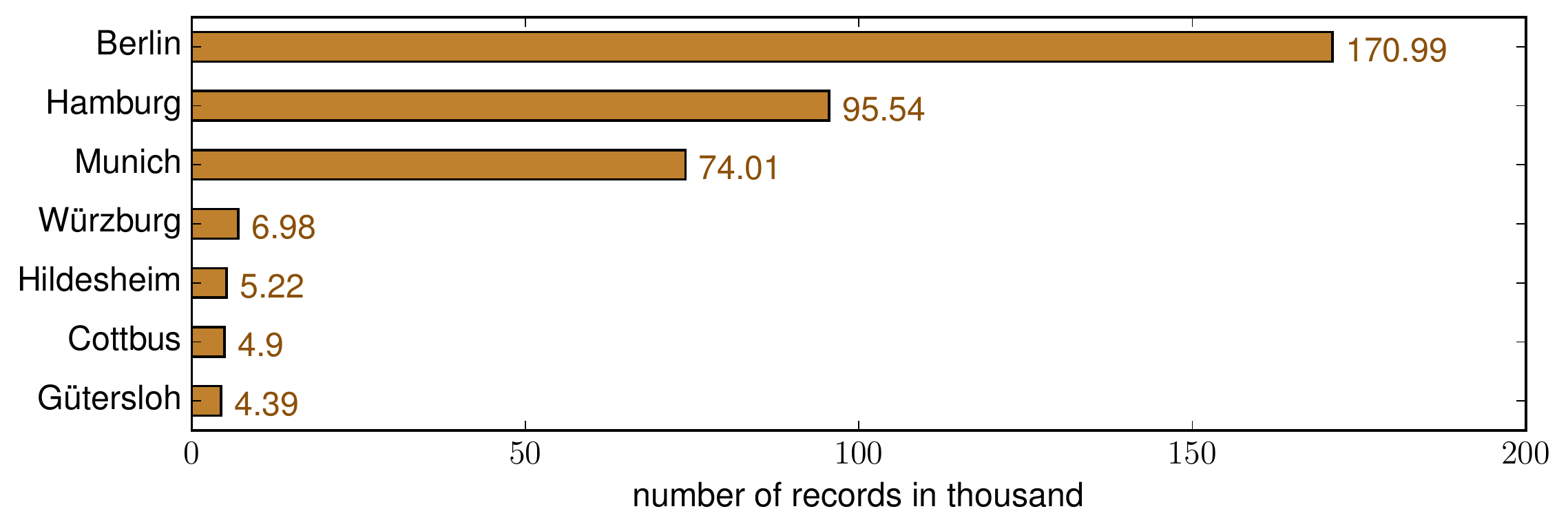}
  \caption{Number of measurements in \cities.}
      \label{fig:citiesNumMeasurements} 
\end{figure}

The number of measurements per cities can exemplarily be seen in Figure~\ref{fig:citiesNumMeasurements}.
In Germany there are 81 cities with more than 100\,000 inhabitants, which are referred to as so-called large cities (\textit{Großstädte}). We looked at the cities individually in the results. In \textbf{Figure~\ref{fig:citiesNumMeasurements}}, we selected seven German cities according to their population: Berlin (3\,644\,826 inhabitants), Hamburg (1\,841\,179 inhabitants), and Munich (1\,471\,508 inhabitants) are the three largest cities, while Hildesheim (101\,990 inhabitants), Cottbus (100\,219 inhabitants), and Gütersloh, (100\,194 inhabitants) are the smallest ones. In addition, we included the medium-sized town Würzburg (127\,880 inhabitants) to our evaluations.
Thus, it can be seen that even for the smallest city Gütersloh 4\,391 measurements were collected.

\section{Evaluation}\label{sec:evaluation}

To analyze the mobile Internet experience from end user perspective, we evaluated the collected data from three different aspects.
First, we focus on all measurements and make statements for Germany in general.
Afterwards, we compare the network performance of the 16 different German states.
Finally, we focus on selected German cities and correlate the results with geographic information.

\subsection{General Figures for Germany}\label{subsec:general_german}

\begin{table*}[t]
    \centering
	\caption{Overall network statistics of Germany.}
	\label{tab:stats_germany}
    \begin{tabular}{@{}lllllllll@{}}
        \toprule
                        & \# measurements   & mean    & std     & min  & max     & 25\% quantile & 50\% quantile    & 75\% quantile \\ 
        \midrule
        dl throughput   & 2\,555\,557           & 15.3   & 12.9   & 0       & 167.1  & 6.5          & 11.4            & 20.1       \\
        ul throughput   & 2\,555\,557           &  7.5   &  6.8   & 0       & 99.9   & 2.4          & 5.1             & 10.8       \\
        latency         & 2\,368\,764           & 27.4   & 44.3   & 0.7    & 4862.8 & 16.9         & 20.6            & 26.6       \\
        jitter          & 2\,370\,697           &  3.9   &  5.0   & 0       & 199.8  & 1.6          &  2.7            &  4.6       \\
        \bottomrule
    \end{tabular}
\end{table*}

Before comparing the network performance of specific regions or states in Germany, we will give an overview of the overall network statistics of Germany based on our dataset.
Thus, \textbf{Table~\ref{tab:stats_germany}} shows basic statistics concerning the download and upload throughput, latency, and jitter.
Due to data privacy reasons, we will refrain from naming the individual providers or publishing statistics for each provider.
Focusing on the measured download throughput, in Germany, a mean of 15.4\,Mbps with a standard deviation of 12.9\,Mbps and a maximum at 167.1~Mbps was measured.
The standard deviation is very high, resulting in a coefficient of variation of 83.9\%, which means a great level of dispersion around the mean. 
The mean is therefore a characteristic with poor informative value. 
Hence, we go further into detail into the quantiles in order to better evaluate the throughput numbers. 
The interquartile range is 6.5\,Mbps up to 20.1\,Mbps, which means that 75\% of all measurement values lie in this range.
1.3\% of the values are below 1\,Mbps while 12.8\% of the values are above 30\,Mbps.

When looking at the measured download throughput values, the question arises which apps were able to be served with high QoE.
According to~\cite{casas2015exploring}, the required throughput is 1\,Mbps for the web-based social media platform Facebook, while 4\,Mbps are required for Google Maps. 
Having a look at video streaming, Netflix, for example, provides the following recommendations for video streaming to avoid stalling: 5\,Mbps for HD and 25\,Mbps for Ultra HD videos\footnote{Internet connection speed recommendation for different video resolutions: https://help.netflix.com/en/node/306 (Accessed: 2020-01-22)}.
Concerning social media apps like Facebook, only in 1.3\% the download throughput measurements was lower than 1\,Mbps.
In 12.6\% of our measurements and thus for every eighth end user, the measured download throughput was lower than 4\,Mbps which probably resulted in bad QoE for using popular apps like Google Maps.
Concerning video streaming, only 17.5\% of all users would have been able to stream Ultra HD videos with high satisfaction.

When looking at the other measured values like upload throughput, latency, and jitter, it can also be seen that the standard deviations are very high, too.
For the upload throughput, the mean value is 7.5\,Mbps with a standard deviation of 6.8\,Mbps, which means a coefficient of variation of 89.7\%.


\begin{figure} [t]
    \centering
    \includegraphics[width=.98\columnwidth]{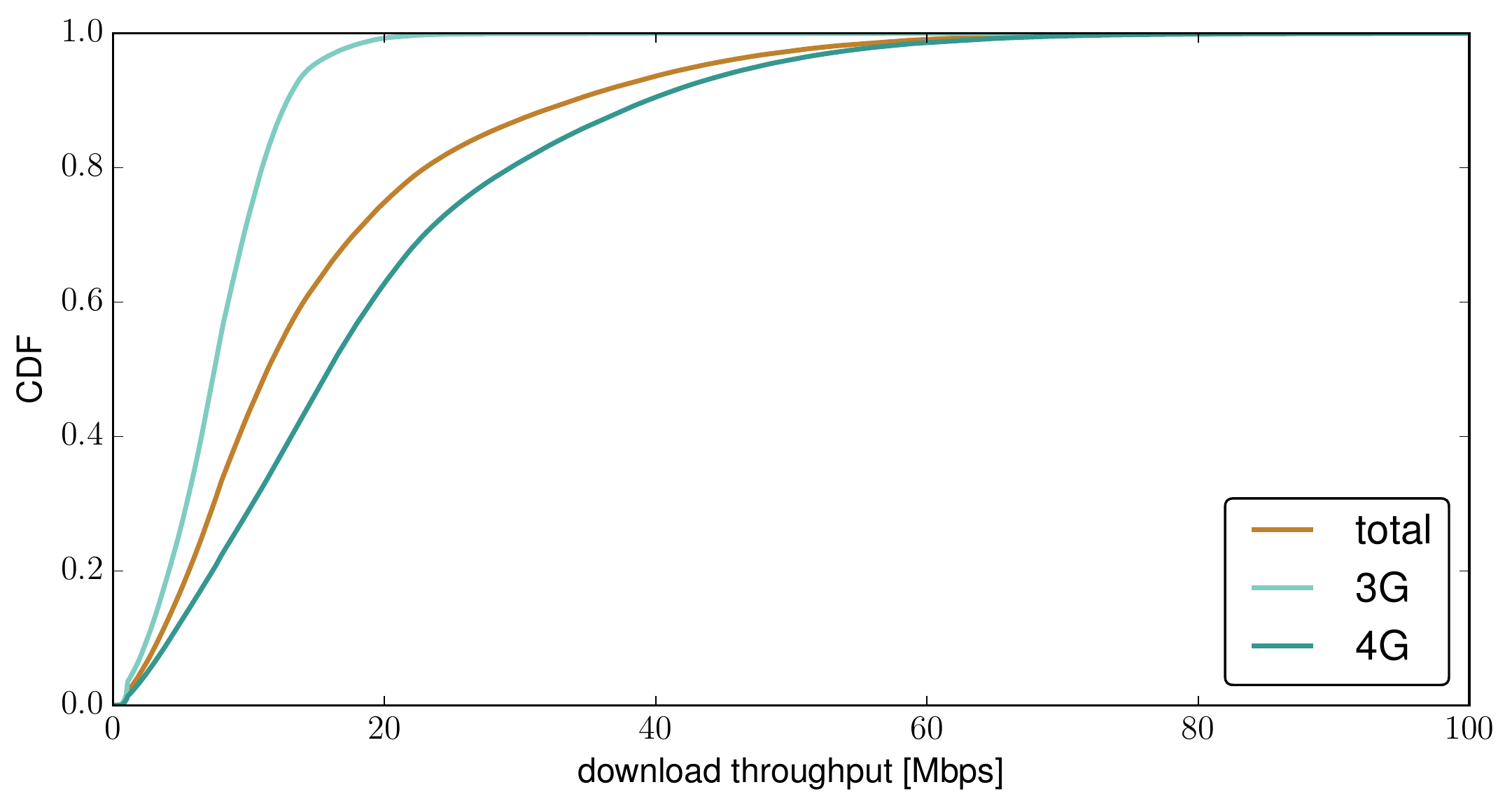}
  \caption{CDF of the download throughput per access technology, cut off at 100\,Mbps.}
      \label{fig:cdf_dl_throughput} 
\end{figure}

Having a look at the used access technology, 
in 66.8\% 4G was used, in 32.9\% 3G, in less than 0.1\% 2G, and in 0.3\% the technology could not be identified.
We assume that the 2G value should be used with caution due to a certain measurement bias.
The measurement method requires smartphones, which means that the proportion of 2G throughput measurements does not correspond to the actual proportion of 2G users in the network (consider the mass of 2G IoT devices), but refers to smartphones with a 2G connection.
\textbf{Figure~\ref{fig:cdf_dl_throughput}} shows the cumulative distribution function (CDF) of the download throughput per access technology as well as for all measurements, cut off in the figure at 100\,Mbps.
For 3G, a mean download throughput of 7.7\,Mbps with a maximum of 77.9\,Mbps and a standard deviation of 4.1\,Mbps was measured.
In comparison, for measurements using 4G, a mean of 19.1\,Mbps with a maximum of 167.1\,Mbps and a standard deviation of 14.0\,Mbps was recorded.
While the mean value is higher here, the coefficient of variation is also significantly higher (73.4\%) compared to 3G (53.3\%).

By comparing the values with the required specifications in the IMT-2000~\cite{IMT-2000} for 3G and IMT-Advanced~\cite{IMT-Advanced} for LTE, the values for 3G are on average above the required throughput values of 2\,Mbps for stationary users. 
The required values for pedestrians and moving vehicles are 384\,kbps and 144\,kbps for vehicles in IMT-2000. 
For 4G, the values are significantly below the peak rates of the 3GPP LTE or LTE Advanced specifications with 327\,Mbps, respectively 1000\,Mbps with low mobility. 
The IMT-Advanced requires a throughput rate in the downlink and uplink of 1 Gbps. 
However, if you calculate an LTE performance with 5\,MHz, 64-QAM, an error correction rate of 5/6 and frame duration of 0.5 ms (200 frames per second), you get 19\,Mbps or about 38\,Mbps for 2x2 MIMO. 
Taking into account that the German mobile providers usually try to deploy 10\,MHz or larger bands, the results show that other factors such as channel quality to the user or the number of users in a cell have a significant influence on the actual throughput rates.


\subsection{German State Comparison}

\begin{figure} [t]
    \centering
    \includegraphics[width=\columnwidth]{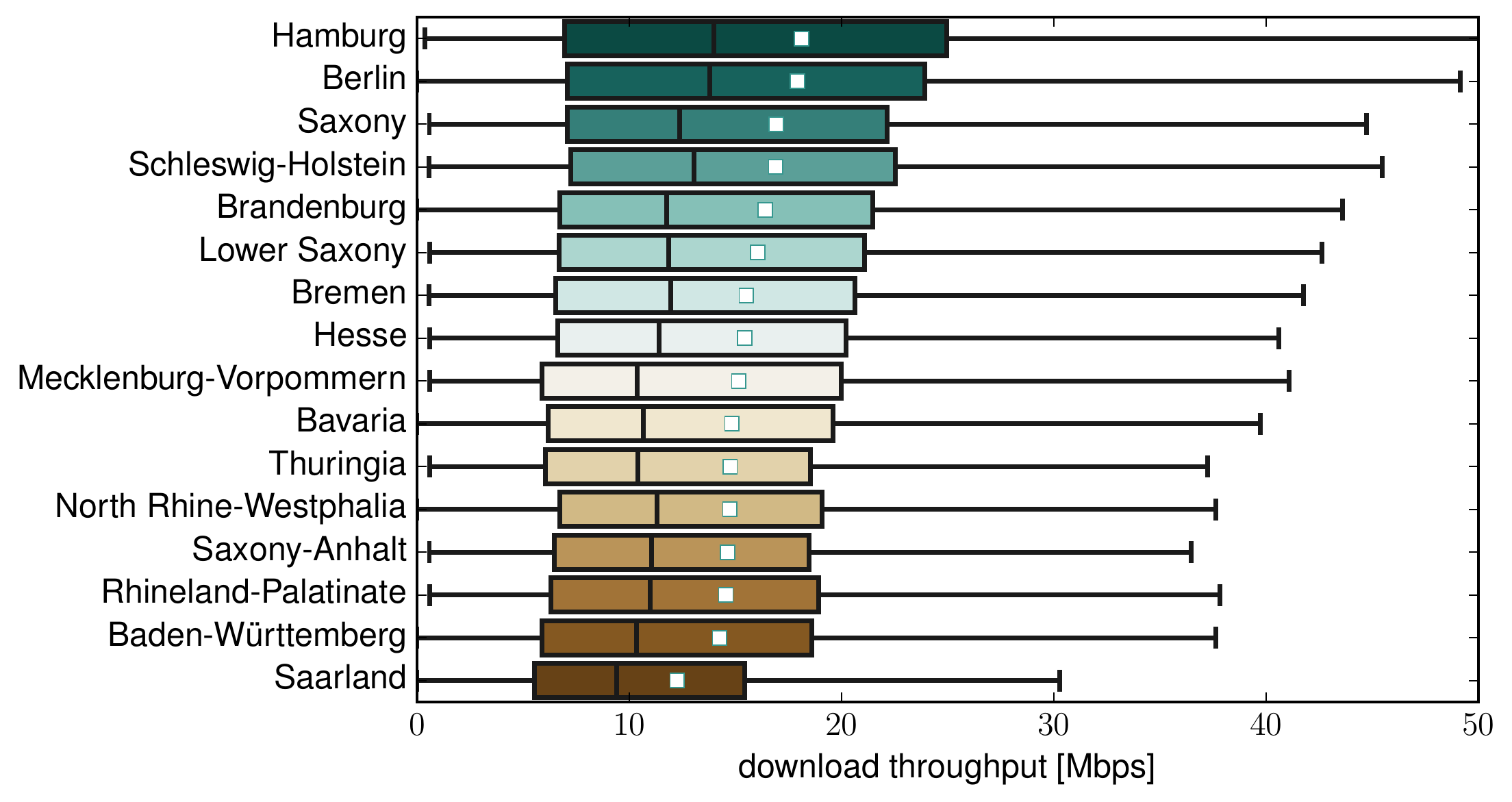}
  \caption{Download throughput per region.}
      \label{fig:regions_dl_throughput} 
\end{figure}

To go more into detail, we investigated the differences between the network conditions in the 16 German states.
\textbf{Figure~\ref{fig:regions_dl_throughput}} shows boxplots of the measured download throughput per state.
The median is marked in each box as black vertical, while the mean is presented as white rectangle. 
6 out of 16 states have a mean download throughput higher than 16\,Mbps.
The highest mean download throughput can be seen in Hamburg (18.1\,Mbps) and Berlin (17.9\,Mbps), lowest in Saarland (12.3\,Mbps).
Nevertheless, Hamburg has the second highest standard deviation of 14.8\,Mbps (headed by Berlin with 14.9\,Mbps), while Saarland with 10.3\,Mbps has the lowest standard deviation.
Thus, it is shown that the mean download throughput differs significantly among all 16 states.

Next, we focus on the upload throughput, where a similar picture can be drawn.
The means of the upload throughput per state show significant differences.
Here, again Hamburg (8.2~\,Mbps) and Berlin (8.2\,Mbps) show the highest values while Saarland (6.6~\,Mbps) the lowest.

Looking at the correlation between the measured network characteristics and statistics of each state, influencing factors can be found.
For the mean download throughput and the population density per state, a significant positive linear relationship (Pearson correlation: $\tau=0.57$, $p{\text-}value=0.02$) is visible.
Thus, the higher the number of inhabitants per km$^2$, the higher the mean download throughput, which shows the deployment strategies of German mobile network providers, probably due to economic reasons.
In addition, we found that the standard deviation of the download throughput per region increases significantly for increasing population density (Pearson correlation: $\tau=0.51$, $p{\text-}value=0.04$). 
For example, Berlin, with a population density of 4\,090 inhabitants per km$^2$, has a standard deviation of 14.9\,Mbps for the download throughput, while Saarland (385 inhabitants per km$^2$) has a standard deviation of 10.3\,Mbps for the download throughput.
Furthermore, a significant correlation (Pearson correlation: $\tau=0.59$, $p{\text-}value=0.01$) concerning the mean upload throughput and the population density was found.
Focusing on the influence factors of jitter, a correlation to the size of the state (Pearson correlation: $\tau=0.59$, $p{\text-}value=0.02$) as well as the population (Pearson correlation: $\tau=0.52$, $p{\text-}value=0.04$) is visible.

To investigate the influence of the used access technologies in each state, \textbf{Table~\ref{tab:states_technologies}} shows the percentage of measurements conducted using 4G, 3G, 2G, or unknown technology.
Again, we assume that the value for 2G should be used with caution due to the measurement methodology, see Section~\ref{subsec:general_german}.
Berlin, Hamburg, and Bremen show the highest number of 4G measurements, while Brandenburg, Thuringia, and Saarland the lowest.
In all states, at most 0.01\% of the measurements were done using a smartphone with 2G access.
By correlating the mean download as well as the mean upload throughput with the percentage of 4G measurements, high positive linear correlations are visible (Pearson correlation download: $\tau=0.80$, $p{\text-}value<0.01$, upload: $\tau=0.79$, $p{\text-}value<0.01$).
This result matches our previous findings that on average measurements conducted using 4G have a significant higher download and upload throughput than using 3G.
Concerning jitter and latency, the used technology showed no influence.

\begin{table}[t]
    \centering
    \caption{Usage of access technologies in the dataset}
    \label{tab:states_technologies}
    \begin{tabular}{@{}lllrl@{}}
        \toprule
                              & 4G    & 3G    & 2G   & unk. \\ 
                              \midrule
        Baden-Württemberg      & 62.2 & 37.4 & $<$0.1 & 0.3   \\
        Bavaria                & 67.6 & 32.1 & $<$0.1 & 0.3   \\
        Berlin                 & 74.8 & 24.8 & $<$0.1 & 0.4   \\
        Brandenburg            & 61.6 & 38.2 & $<$0.1 & 0.2   \\
        Bremen                 & 69.2 & 30.7 & $<$0.1 & 0.1   \\
        Hamburg                & 74.5 & 25.2 & $<$0.1 & 0.3   \\
        Hesse                  & 65.4 & 34.2 & $<$0.1 & 0.4   \\
        Lower Saxony           & 66.8 & 32.9 & $<$0.1 & 0.4   \\
        Meck.-Vorpom.          & 63.0 & 36.8 & $<$0.1 & 0.2   \\
        North Rhine-Westph.   & 66.9 & 32.7 & $<$0.0 & 0.3   \\
        Rhineland-Palatinate   & 63.6 & 36.1 & $<$0.1 & 0.3   \\
        Saarland               & 59.0 & 40.6 & $<$0.1 & 0.3   \\
        Saxony                 & 68.2 & 31.3 & $<$0.1 & 0.5   \\
        Saxony-Anhalt          & 66.2 & 33.4 & $<$0.1 & 0.4   \\
        Schleswig-Holstein     & 68.7 & 31.1 & $<$0.1 & 0.3   \\
        Thuringia              & 60.9 & 38.6 & $<$0.1 & 0.4   \\ 
        \bottomrule
    \end{tabular}
\end{table}


\subsection{German City Comparison}

\begin{figure} 
    \centering
    \includegraphics[width=.98\columnwidth]{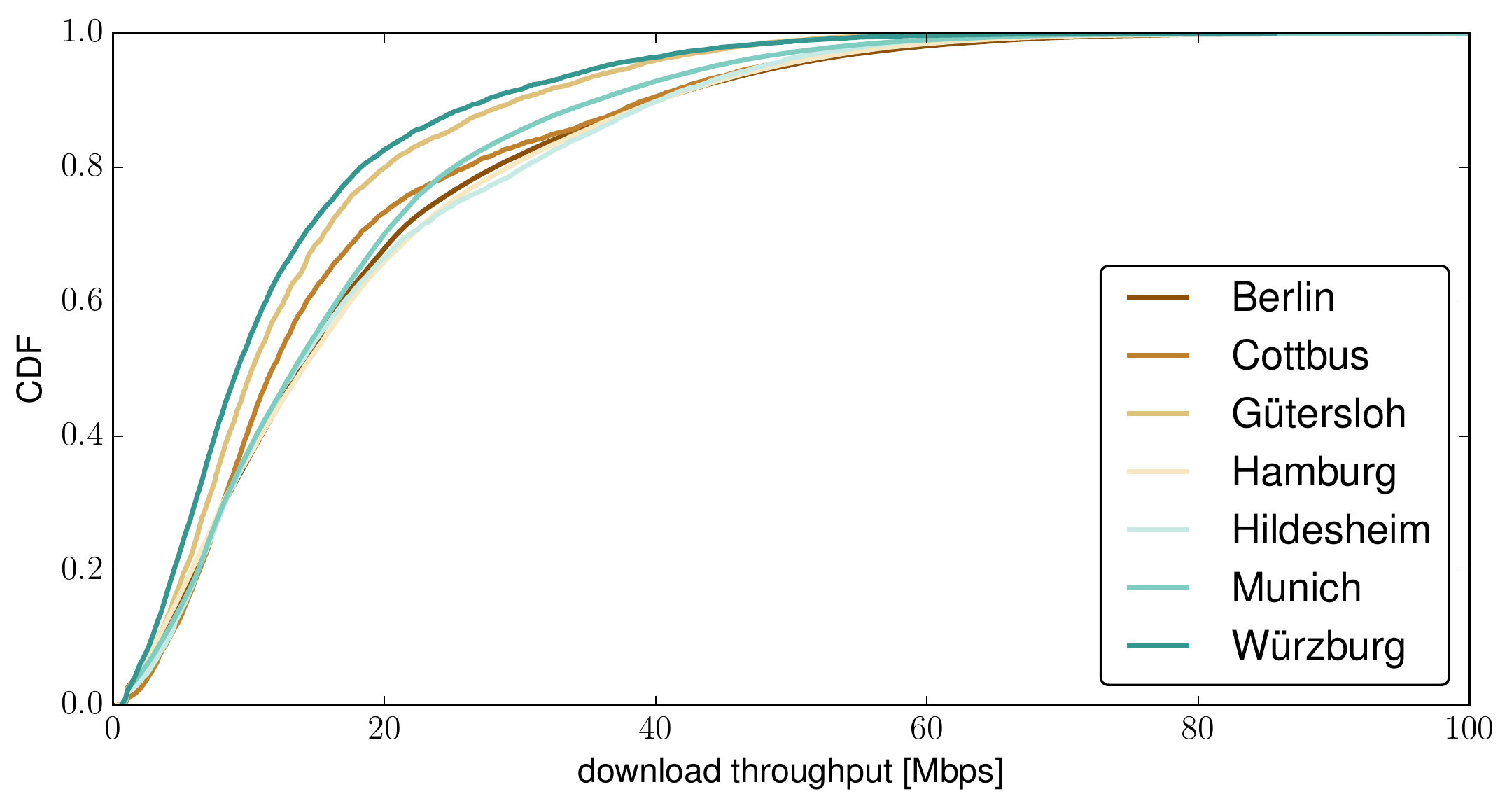}
  \caption{Download throughput per city for \cities.}
      \label{fig:cities_dl_throughput} 
\end{figure}

Finally, we will have a look at the measurement results broken down to individual cities.
Starting again with the download throughput, \textbf{Figure~\ref{fig:cities_dl_throughput}} shows the cumulative distribution of the measured values for seven selected cities.
Here, Hildesheim and Hamburg show the highest mean download throughput with 18.2\,Mbps and 18.1\,Mbps, while Gütersloh and Würzburg show the lowest means (13.7\,Mbps and 12.5\,Mbps).
Nevertheless, Gütersloh and Würzburg show the lowest standard deviations (11.1\,Mbps and 10.9\,Mbps), while Berlin and Hamburg show the highest (14.9\,Mbps and 14.8\,Mbps).
In general, cities with a high mean throughput must naturally also have a higher variation in values.

Having a look at all 81 German large cities (\textit{Großstädte}), we measured the highest mean download throughput in Potsdam (21.3\,Mbps) and Braunschweig (20.1\,Mbps) and the lowest in Reutlingen (10.8\,Mbps) and Bremerhaven (10.9\,Mbps).
Having a look at the standard deviation, Saarbrücken (9.9\,Mbps) and Reutlingen (10.1\,Mbps) showed the lowest values while the highest variations were found in Dresden (15.8\,Mbps) and Potsdam (15.4\,Mbps).

Putting the download throughput in relation to the size of the city, a weak linear correlation (Pearson correlation: $\tau=0.33$ $p{\text-}value\leq0.01$) is visible.
Furthermore, also the standard deviation of the download throughput is influenced by the size of the city (Pearson correlation: $\tau=0.40$ $p{\text-}value\leq0.01$) as well as by the number of inhabitants (Pearson correlation: $\tau=0.32$ $p{\text-}value\leq0.01$).
This correlation could be caused by the fact that larger cities have a high number of inhabitants and thus, the economic benefit for network operators is higher and therefore provides an economical incentive to upgrade the network.
This result is supported by the finding that number of inhabitants as well as the size of the city shows a significant linear correlation to the percentage of 4G measurements (Pearson correlation population: $\tau=0.34$ $p{\text-}value\leq0.01$, size: $\tau=0.33$ $p{\text-}value\leq0.01$).

In general, for all of the 81 large cities, the mean download and upload throughput increases by increasing percentage of 4G measurements.
Here, a significant correlation of $\tau=0.64$ ($p{\text-}value\leq0.01$) for download, and a correlation of $\tau=0.69$ ($p{\text-}value\leq0.01$) for upload throughput is visible.
In this case, jitter and latency decrease.


\section{Conclusion}\label{sec:conclusion}

In this paper, the mobile Internet experience for Germany was evaluated with the help of crowdsourcing from the perspective of an end user. Based on a dataset from Tutela Ltd. with 2.5 million throughput measurements from January to July 2019, statistics for various German areas were created and examined. Insights into this emerging methodology of crowdsourced network measurements were given and the advantages of this method were highlighted. The paper contains statistics and conclusions for several large cities and regions in Germany compared to general statements for Germany. One contribution of the paper is to show how the measured values differ between individual measurements and the mean value for an area. 

The results present a heterogeneous measurement result. Compared to the figures throughout Germany, high variances in the measurements are already visible for the states. This is mainly due to the deployment status of the network, which is being driven forward in areas with higher population numbers due to economic factors. Furthermore, higher throughput values are to be seen in areas with high population density. 
For Germany, the interquartile range is 6.5\,Mbps up to 20.1\,Mbps, which means that 75\% of all measurement values lie in this range.
1.3\% of the users got less then 1\,Mbps throughput which is below the recommended value to use web-based apps such as Facebook according to~\cite{casas2015exploring}. For every eighth end user, the measured download throughput was lower than 4\,Mbps which probably resulted in bad QoE when using popular apps like Google Maps. Only 17.5\% of all users would have been able to stream Ultra HD videos with high satisfaction. The 3G throughput for Germany according to our dataset lies at 7.7\,Mbps. The 4G throughput was measured at 19.1\,Mbps. While the mean value is higher here, the coefficient of variation is also significantly higher (73.4\%) compared to 3G (53.3\%). Going into detail about the cities, the differences in the throughput are even more severe. Hamburg has the highest average throughput with 18.1\,Mbps, while Saarland is at the bottom of the list with 12.3 Mbps. It is shown that the mean download throughput differs significantly among all 16 states.
 
Overall, it can be concluded that, firstly, individual views are necessary and, secondly, different regions have to be considered in order to obtain a holistic picture of the status of the mobile network performance in Germany. The throughput is higher in densely populated areas such as large cities than in smaller cities. Overall, mean downlink throughput and city size correlate, and standard deviation of downlink throughput and city size correlate. Large cities are therefore better developed, but also fluctuate more in throughput.

\bibliographystyle{IEEEtran}
\bibliography{references}
\balance

\end{document}